\documentclass[1p,authoryear]{elsarticle}

\usepackage{enumerate}
\usepackage{maltese}
\usepackage[table]{xcolor}
\usepackage{rotating}
\usepackage{setspace} 
\usepackage{url}
\usepackage{float}

\author[1,2]{Joseph Caruana\corref{cor1}}
\author[3]{John Wood}
\author[4]{Erica Nocerino}
\author[5]{Fabio Menna}
\author[6,7]{Aaron Micallef}
\author[3]{Timmy Gambin}

\cortext[cor1]{Corresponding author, joseph.caruana@um.edu.mt}

\address[1]{Department of Physics, Faculty of Science, University of Malta, Msida MSD 2080, Malta}
\address[2]{Institute of Space Sciences \& Astronomy, University of Malta, Msida MSD 2080, Malta}
\address[3]{Department of Classics and Archaeology, Faculty of Arts, University of Malta, MSD 2080 Msida, Malta}
\address[4]{Dipartimento di Scienze Umanistiche e Sociali, Universit\`{a} degli Studi di Sassari, Via Roma 151, 07100 Sassari (SS), Italy}
\address[5]{3D Optical Metrology (3DOM) Unit, Bruno Kessler Foundation (FBK), Trento, Italy}
\address[6]{Department of Geosciences, University of Malta, Msida MSD 2080, Malta}
\address[7]{GEOMAR Helmholtz Centre for Ocean Research, Kiel, Wischhofstrasse 1-3, D-24148 Kiel, Germany}

\begin{document}

\begin{abstract}
The Azure Window was a natural arch situated in the west coast of Gozo (Maltese Archipelago) that collapsed in March 2017. We employ a Diver Propulsion Vehicle-mounted camera system to capture data for the 3D-modelling of this collapsed arch via photogrammetry. We demonstrate use of this method to document complex underwater geomorphology spread across a large area, and draw up a geomorphic assessment of the site and collapse event on the basis of this 3D-model. The methodology enables a reconstruction and understanding of the collapse event. On account of the high-resolution attained, we are able to cross-match the principal submerged components with their pre-collapse location; this enables an understanding of the dynamics of the collapse, confirmation of rock break-up along existing joints, and mapping the distribution of the rock and debris from the collapse event. We conclude that the key stages in the collapse of the Azure Window entailed erosion at the base of the pillar, leading to the latter's collapse in the southwest direction, breaking into two main sections that separated along the lithological boundary. We also find clear evidence that separation of some sections of the pillar followed pre-existing joints. The bridge collapsed vertically upon loss of support from the pillar, breaking into two main components and many other fragments. We also document further changes at the site post-collapse. We show that this approach can be utilised to understand and characterise such events even when significant time has elapsed since collapse, and rocks have already undergone erosion and significant marine growth.
\end{abstract}

\begin{keyword}
coastal arch - arch collapse - underwater photogrammetry - Azure Window
\end{keyword}

\title{Reconstruction of the collapse of the `Azure Window' natural arch via photogrammetry.}

\maketitle

\section{Introduction}

Coastal arches are the product of erosion. The mechanical and hydraulic action of waves can result in the formation of a range of landforms, such as marine notches, shore platforms (e.g.~\citealp{trenhaile2000}; \citealp{trenhaile2001}), marine caves, and arches. The standard description of arch formation draws a picture of preferential erosion acting on a weaker or softer basal layer of the headland, carving out a recess that gradually progresses to form a cavern or arch (see, e.g., \citealp[p.~281]{johnson1919}; \citealp[pp.~190-193]{sunamura1992}; \citealp{shepard1983}; \citealp{trenhaile1987}; \citealp{clark1995}; \citealp{trenhaile1998}; \citealp[pp.~515-522]{duff1998}; \citealp[pp.~153-157]{woodroffe2002};  \citealp{may2019}). The progression of this erosion sequence may be abetted by the existence of weak points and discontinuities in the rock such as existing joints or fractures, bedding, and cross-bedding (\citealp{wilson1952}; \citealp{wood1968}; \citealp{stevens1988}; \citealp[pp.~188-190]{sunamura1992} and references therein; \citealp{cruikshank1994, trenhaile1998, clark1995, dickson2004, grab2011}). These factors, as well as rock stresses \citep{bruthans2014, bruthans2017, rihosek2019, filippi2018}, play an important role in the development of a natural rock arch, whose geometry can span a range of shapes bounded by two limiting cases: an inverted catenary on the one hand, and a comparatively less stable beam on the other; while an arch may seek to minimise tensile stresses, the aforementioned discontinuities can be a controlling factor leading to the arch assuming a form that compromises long-term stability \citep{moore2020}.

The final stage in the evolution of coastal arches is their collapse, a recent well-known example of such an event being the demise of Darwin's Arch in the Galapagos in May 2021. The collapse tends to leave the headland on one side and a stack on the other. Not all stacks, however, are the residual of arch-collapse; some, for example, are the product of erosion of weaker areas of the headland, resulting in the excavation of a transverse inlet with more resistant rock remaining in place (e.g., \citealp[pp.~59-61]{bird2000}). Stacks are found at many sites around the world (see, e.g., \citealp{bird2010} for numerous examples), and a number of them have been the subject of documentation or study. Such examples include: (i)~La Jolla, California, where a number of arches and stacks visible in old photographs, many of which have since collapsed, were analysed (\citealp{shepard1983}, \citealp[pp.~274-275]{kuhn1983}); (ii)~Table Head, Glace Bay, Cape Breton Island, Canada, where historical photographs were used to trace the history of two stacks which followed the collapse of an arch that had formed via the convergence of two caves cut into two sides of a promontory (\citealp[pp.~150-151]{shepard1983}, following \citealp{johnson1925}); (iii)~The series of three chalk stacks known as `The Needles' at the Isle of Wight, England, which were dissected through chalk (see \citealp[p.~193]{sunamura1992} and references therein); (iv)~Chalk stacks at Ballard Down, England \citep{may1985}, being the result of erosion occurring in vertical joints; (v)~Taito-misaki, Japan, where the stack was not the end-result of arch-collapse, but a fault lying between the stack and the mainland that extended to the very cliff-top (\citealp[pp.~191-193]{sunamura1992}; \citealp{sunamura1973}); (vi)~Baja California, Mexico, where stack formation resulted from the intersection of two keyholes \citep{clark1995}; (vii)~The island of Hongdo off the southwestern Korean Peninsula, with quartzite stacks (along with arches and caves) being photographically documented and related to parting planes in the rock \citep{johnson2013}.

However, explanations of stack formation are largely of a descriptive or phenomenological nature. Despite their being a common landform, only a small number of studies have focussed on their genesis and evolution, such that their dynamics remain poorly constrained (see, e.g., \citealp{naylor2010, naylor2014}). Efforts to address this lacuna are represented by a study of the stacks of Hopewell Rocks, New Brunswick, Canada, most of them resulting from dissection of well-developed joint planes \citep{trenhaile1998}, more recent exploratory modelling investigating the role that abrasion plays in their evolution \citep{limber2015}, and an investigation of their longevity, focussing on the Twelve Apostles in Victoria, Australia \citep{bezore2016}.

Events of arch collapse and resultant stack formation are hard to observe in real-time, given that the collapse happens quickly and is likely to occur during rough weather, with a smaller probability of witnesses. Moreover, in the case of coastal arches, the resulting submerged deposits are generally difficult to map, especially if located in shallow waters, for it is difficult to employ standard seafloor mapping techniques, such as the deployment of a vessel equipped with a multibeam echosounder. Addressing this gap in knowledge has important implications both for improving our understanding of the evolution of coastal arches and in terms of hazard assessment, particularly in relation to predicting when and how a coastal arch might collapse.

Situated at the site of Dwejra on the island of Gozo (Maltese archipelago), the Azure Window, known in Maltese as \textit{It-Tieqa tad-Dwejra}, was an internationally-known natural arch that collapsed on the 8th of March, 2017. In this work, we carry out an investigation of the dynamics of the collapse via comprehensive underwater photogrammetry of the debris on the seafloor. This, in turn, enables us to gain further insight into the collapse event. Specifically, the aims of this paper are to:

\begin{enumerate}
\item Demonstrate the viability of a specially developed setup whereby a camera system mounted on a continuously moving Diver Propulsion Vehicle (DPV) is utilised for surveying submerged geomorphology.

\item Employ photogrammetry to construct a 3D-model for an underwater site that is spread over a very large area, overcoming numerous challenges.

\item Present the first geomorphic assessment of the remains of the Azure Window natural arch.

\item Reconstruct the key stages of the natural arch collapse event.

\end{enumerate}


\section{Study Area}\label{sec:studyarea}

\subsection{Geological Setting}\label{sec:geosetting}

The stratigraphy of the Maltese Islands was first outlined in \cite{spratt1843, spratt1852}. The Maltese Islands are comprised of four main shallow water, Oligo-Miocene sedimentary formations \citep{pedley1976}. From base to top, these formations include: (i) Lower Coralline Limestone ($\dot{Z}$\textit{onqor}, maximum thickness of 140~m) - thickly bedded and coarse bioclastic limestones, biomicrites and massive coralline algal limestones deposited in a shallow marine environment; (ii) Globigerina Limestone (\textit{Franka}, maximum thickness of 200~m) - taking its name from the planktonic Foraminifera Globigerina, it consists of biomicritic packstones, wackestones, and marls deposited in outer shelf environments; (iii) Blue Clay (\textit{Tafal}, maximum thickness of 75~m) - banded plastic kaolinitic marls and clays, generally with a variety of fossil macrofauna, which were deposited in water depths in excess of 200~m; (iv) Upper Coralline Limestone (\textit{Qawwi ta' Fuq}, thickness of more than 100~m) - coarse grained wackestones, packstones and mudstones rich in coralline algae, molluscs and echinoids deposited in a shallow water environment.\footnote{The principal difference in Spratt to the modern-day description lies in the nomenclature that he used, namely (1) Coral Limestone, (2) Yellow sandstone and blue clay, (3) Freestone, and (4) Semi-crystalline limestone.} Some regions also exhibit thin Quaternary deposits (the earliest descriptions of which are found in \citealp{spratt1867, trechmann1938}).

The Maltese Islands are intersected by two normal fault systems \citep{illies1981}. The older system comprises NE-SW trending faults that were active from the Early Miocene to mid-Pliocene times \citep{pedley1990, gardiner1995}. The younger generation of faults are oriented NW-SE, with their formation or re-activation taking place in the late Pliocene to Quaternary \citep{gardiner1993}.

\begin{figure*}
  \includegraphics[width=\linewidth]{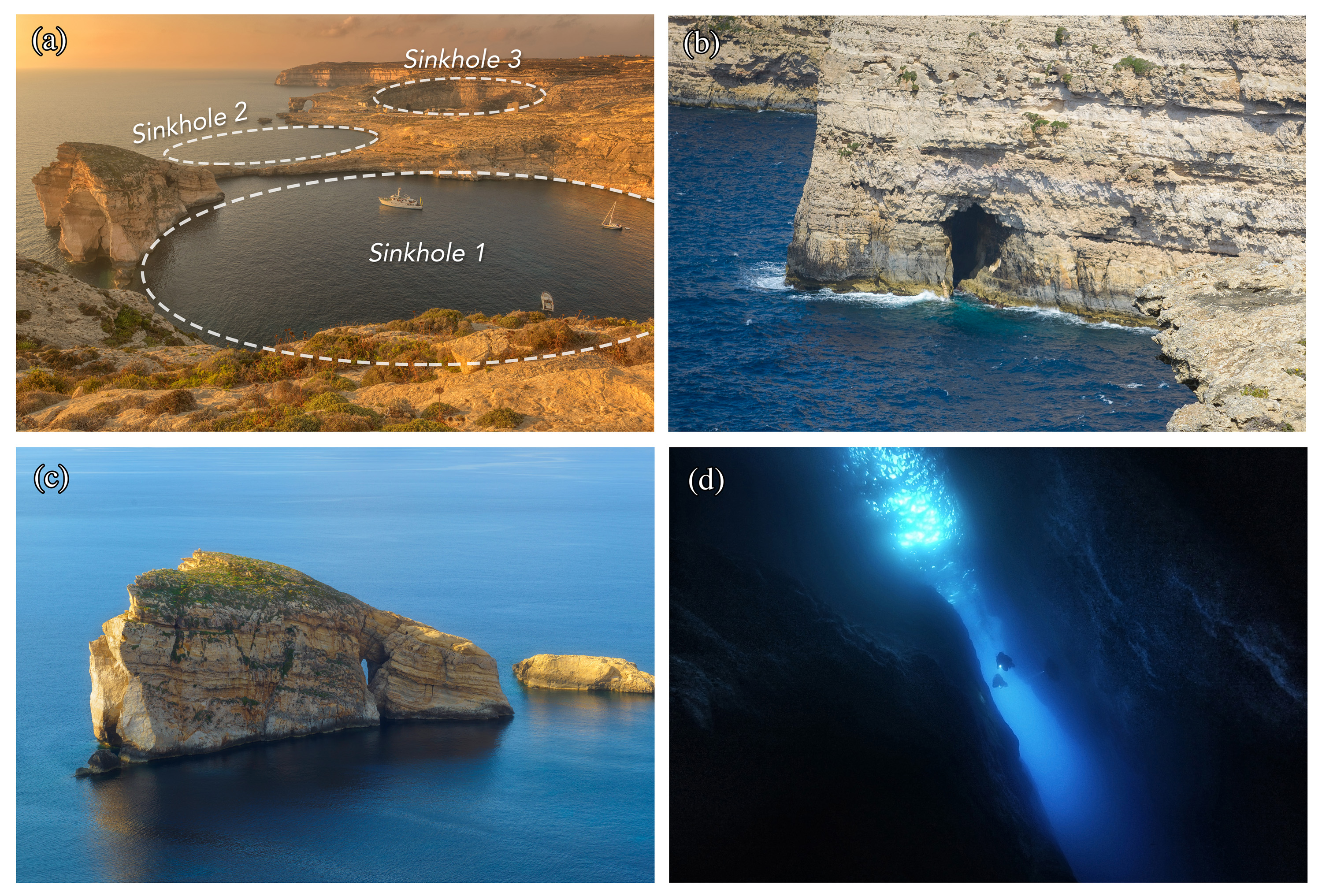}
  \caption{Some of the geomorphic features visible at Dwejra. (a) Three of the sinkholes at Dwejra, where Sinkhole 1 is known as \textit{Il-Qala tad-Dwejra}, Sinkhole 2 is located at \textit{Il-Qasir}, and Sinkhole 3 is \textit{Il-Qawra} (or the `Inland Sea'). Note the Azure Window to the west of Sinkhole 3. (b) A sea cave at the cliffs of Dwejra. (c) The stack known as `Fungus Rock' (\textit{Il-\.{G}ebla tal-\.{G}eneral}). (d) Underwater view of the Inland Sea tunnel (\textit{L-G$\hbar$ar tal-Qawra}).  (Images are copyrighted by J.~Caruana.)}
  \label{fig:geomorphological_features}
\end{figure*}

\begin{figure*}
  \includegraphics[width=\linewidth]{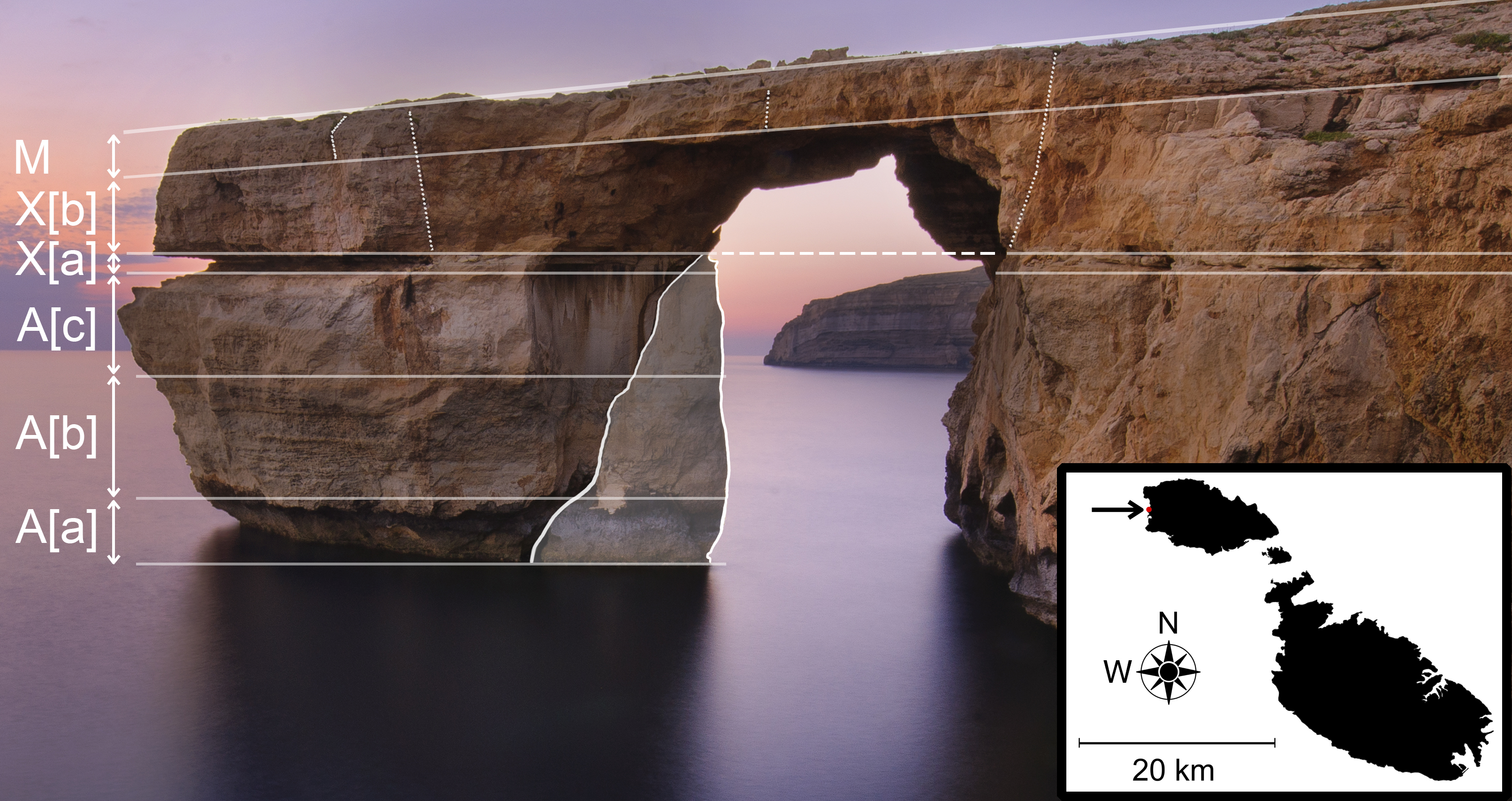}
  \caption{Lithology of the Azure Window, after \cite{pedley1978}. Labels `A', `X', and `M' denote Attard, Xlendi and Mara members respectively. Letters in square brackets denote principal subdivisions of facies. Dotted lines denote visible cracks on the Xlendi and Mara members. The horizontal, dotted line underneath the bridge marks the boundary above which rockfall (largely from the X[b] member) took place since a \textit{c.}~1900 photo by Richard Ellis (Panel (a) of Fig.~\ref{fig:arch_progress}). The shaded region on the pillar marks a large section that collapsed in 2012. The inset shows a map of the Maltese Islands with the location of the former Azure Window (west coast of Gozo), marked with an arrow. The site coordinates are 36$^{\circ}$03$^{\prime}$12.9$^{\prime \prime}$N, 14$^{\circ}$11$^{\prime}$19.1$^{\prime \prime}$E. (Image copyright: J.~Caruana.)}
  \label{fig:lithology_annotated}
\end{figure*}

The Azure Window was located in Dwejra, a coastal site in the northwestern part of the island of Gozo that exhibits a wide range of geomorphic features of interest formed in Globigerina and Lower Coralline Limestone formations (see Fig.~\ref{fig:geomorphological_features}). These include a high concentration of solution subsidence structures\footnote{These were also first described by Spratt, who used the term `depressed basin'.} possibly arising from the dissolution of sub-surface evaporite salts (\citealp{gatt2013}, \citealp{galve2015}, \citealp{gatt2019}), their evolution being modulated by erosion \citep{soldati2013}, as well as karstic conduits, natural arches, sea caves, underwater caves, tunnels, stacks, vertical cliffs, valleys, and tidal and roof notches (e.g.~\citealp{furlani2017, calleja2019}). The site of the Azure Window was at the southern end of the cliff section and west of one of the largest solution subsidence structures (known as the `Inland Sea'), within the upper section of the Lower Coralline Limestone. This layer is subdivided by \cite{pedley1978} into four members. Named after the places where they are most visibly exposed, from bottom to top these are: Mag\malteseh laq, Attard (A), Xlendi (X), and Il-Mara (M). The latter three of these members (with further subdivision of facies) can be seen in Fig.~\ref{fig:lithology_annotated}. The Attard member comprises horizontally-bedded, shallow marine packestones and wackestones with algal rhodoliths, and large and widely-spaced vertical jointing. The Xlendi member consists of horizontal to cross-bedded lime sand (packstone to grainstone) with coralline algal fragments, and poorly-defined and closely-spaced vertical joints. The Mara member (packstone to wackestone) is poorly developed and exhibits more closely-spaced jointing which makes it prone to small-size rock fragmentation. This member is an echinoid-rich (e.g.~\textit{Scutella subrotunda}) fossil bed. In the case of the Azure Window, the Xlendi and Mara members together constituted a thickness of about 7-8~m, and the Attard member 20~m.

\subsection{The Azure Window: genesis, history, and evolution}

The formation process of the Azure Window would have started with erosion via hydraulic action at sea-level, eventually leading to widening of vertical joints. Cracks widened sufficiently for successive slab-failure to occur, resulting in an arch structure connected to the headland.

\begin{figure*}
  \includegraphics[width=\linewidth]{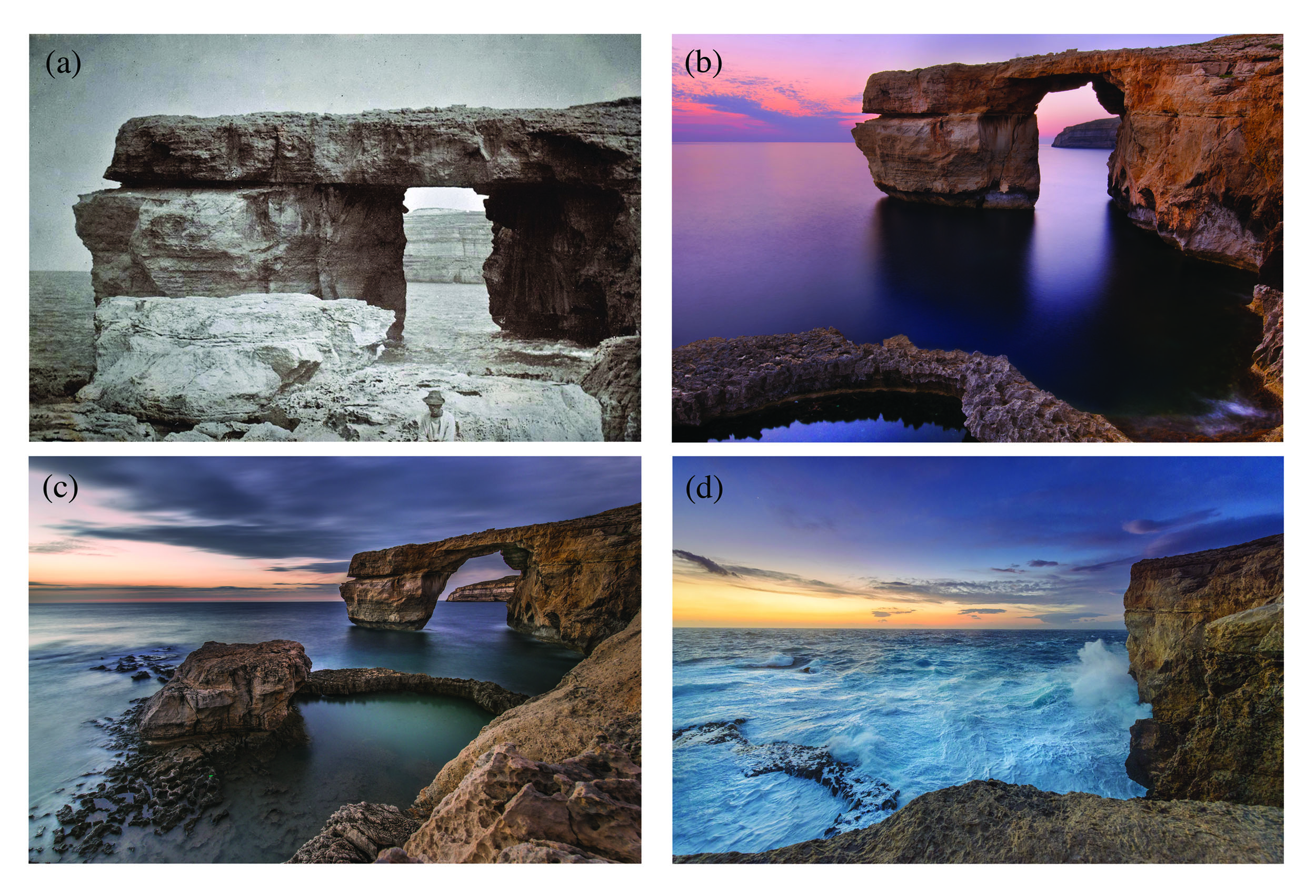}
  \caption{This sequence of images shows the deterioration of the arch leading up to its collapse. Panel (a) is a photograph by Richard Ellis from \textit{c.}~1900, in which the arch exhibits a rectangular shape (practically identical to that visible in the 1981 film `Clash of the Titans'). Panel (b) is an April 2011 photograph that shows substantial thinning of the arch's ceiling. Also visible is a long crack in the pillar that had developed by this time. Panel (c) is an October 2012 photograph in which one can observe a wider arch due to the collapse of a large segment of the pillar, which detached along the crack visible in panel (b). Panel (d) is a photograph of the site on the evening of the arch's collapse on the 8th of March, 2017. Although the weather conditions were worse at the time of collapse, significant wave action can still be seen. Photo (a) is being used with the permission of Giovanni Bonello. Photos (b)-(d) are copyrighted by J.~Caruana.}
  \label{fig:arch_progress}
\end{figure*}

The time of formation of the arch is not known with any precision; at the time of writing, the earliest photographic evidence is found in an 1879 album which carries a photograph by local documentary photographer Richard Ellis \citep{carabott2017}. A similar photo by the same photographer \citep[pp.~87, 100]{ellis2011} dating from \textit{c.}~1900 \citep[p.~106]{bonello2007} is shown in Panel (a) of Fig.~\ref{fig:arch_progress}. It is also depicted in an 1824 painting \citep{ganado2017}, albeit the perspective from which it is drawn does not allow for a definitive conclusion about whether it had already been fully eroded into an open arch at this stage, or whether it was still in its cave phase. 

Neither of the major historians $\dot{\rm G}$an Fran$\dot{\rm g}$isk Abela (1582-1655) and $\dot{\rm G}$an Piet Fran$\dot{\rm g}$isk Agius de Soldanis (1712-1770) make reference to the arch in their works. Since they both mention the nearby Inland Sea tunnel, one might expect that they would have been compelled to refer to the arch as well, had it existed during their lifetime. That being said, famed artist Edward Lear does not seem to have paid any attention to it when he visited the site in 1866 (only 13 years prior to Ellis' photograph); having painted the well-known Fungus rock (a photo of which is shown in Fig.~\ref{fig:geomorphological_features}c), he `toiled up and up' but described the scenery of the coast as `not being nearly as fine as that of Malta' \citep[17th March, 1866 diary entry in][]{lear1866} with no reference whatsoever being made to the Azure Window. It is unlikely that he would have missed to notice such a large geomorphic feature. Absence of reference to the arch need not imply absence of its existence. However, unless any further evidence turns up, at present, the best we can say is that it had already formed (or nearly formed) 193 years prior to its collapse in 2017.

Due to its popularity, there exists ample photographic and video material of the Azure Window from the past decades. Comparison of the Ellis photographs (both the one in the 1879 album and the \textit{c.}~1900 photo shown in this paper) with the arch's depiction in the 1981 film `Clash of the Titans' shows a very similar - in fact, practically identical - shape. Its deterioration progressed quite quickly in the (approximately) 35 years leading to its demise. From a very rectangular form (Fig.~\ref{fig:arch_progress}a) it progressed to an arched aspect due to bridge-thinning (Fig.~\ref{fig:arch_progress}b), losing a significant segment from the pillar in 2012 (Fig.~\ref{fig:arch_progress}c), and collapsing 5 years later (Fig.~\ref{fig:arch_progress}d).

The collapse of the Azure Window occurred on the 8th of March, 2017 on a day characterised by very strong northwesterly winds and large wave-height. There have been no reports of visual witnesses to the collapse event itself. 

Following the collapse of an arch, it is common for a sea-stack to be left behind, an example of which can be observed at the site of Dwejra itself in the form of the aforementioned `Fungus rock' (\textit{\.{G}ebla tal-\.{G}eneral}). In the case of the Azure Window, however, there are no remains above the waterline, with all of its fragmented components now lying submerged at depths ranging between just below the surface down to a depth of $\approx60$~m. 

Previous, relatively recent work focussing on the geology and geomorphology of the Azure Window include: a geotechnical report of the arch prior to its demise \citep{gatt2013}; a preliminary assessment of the post-collapse submerged components, which was carried out for the Dwejra Steering Committee by the lead author of the present paper following a dive conducted on the morning of the 13th of March, 2017, i.e.~5 days following the collapse, when sea conditions allowed for safe exploration 
(unpublished but reported upon in various media, e.g. \citealp{timesofmalta2017, tvm2017, sanford2017, maltatoday2018}, with three of the author's images having been reproduced in \citealp{satariano2019}); and a report on the seismic signature of the collapse event \citep{galea2018}. Since the collapse, numerous photographs and videos have been uploaded by users to online video-viewing platforms. However, the geometry and layout of the arch's underwater remains have not as yet been described, measured, and presented in any published work. 

\section{Methodology}\label{sec:photogrammetry}

\subsection{Principal challenges}
The 3D-models that provide the basis for this study were built via photogrammetry. In the case of the pre-collapse Azure Window, we used drone footage to construct the 3D-model, with scaling being carried out via on-site tape measurements. For the acquisition of underwater images of the post-collapse components, we employed a purpose-built photography setup. Alternative approaches, namely a multibeam survey or the use of a remotely operated underwater vehicle (ROV) were deemed impractical. Multibeam data would deliver accurate bathymetric information that could be processed to produce a digital elevation model (DEM). If such a resulting model were compared to a pre-collapse multibeam survey, one might be able to see topographic changes of the seabed caused by the collapse. However, there are two potential problems: (a) smaller changes might not be recognised, and (b) individual boulders/pieces of the window would not be identified solely through a DEM (whereas photogrammetry delivers both a DEM and unambiguous visual information that allows detailed assessment of specific features).  A small ROV could have been deployed from the coast, but in our case this would have presented some difficulties on account of the local geography/terrain. The ROV would also have lost communication with the USBL transceiver when positioned behind (i.e.~on the other side of) large underwater features. In addition, accurately positioning/locating the ROV underwater would also have been very challenging, as the system is dependent upon GPS; since we were operating in the shadow of a cliff, adequate GPS reception would have been problematic. Operating a ROV from the cliff would also have presented a major obstacle, for the umbilical could have snagged. Moreover, tether management would have been somewhat challenging in either scenario. Besides these considerations, the camera and light quality on small ROVs would not produce the same resolution obtainable by diver-held systems. Deploying a large ROV from a survey vessel was discounted for two reasons: cost and manoeuvrability. A large ROV with a tether management system would be too cumbersome to use on this specific site.

Despite the diving team's extensive prior experience with underwater data acquisition and 3D-model generation, this specific site presented a number of new challenges, both of a logistical and technical nature. In particular:

\begin{enumerate}
\item The size of the site is very large, the total area to be surveyed comprising $\approx$8,000~m$^2$.
\item The site varies greatly in depth, from just below sea level, requiring calm days with flat surface conditions, down to $\approx$60~m.
\item As a consequence of this depth variation, the site exhibits a very wide range of natural (ambient) light. Consequently, producing a consistent set of images requires widely varying exposure and white balance settings.
\item The prevailing wind direction being northwesterly limited the days suitable for diving and data collection.
\item It was not practical to use the closest and most common entry/exit point to the site (the `Blue Hole') as a large amount of equipment would have had to be transported over difficult terrain. The other available and easy access point (the Inland Sea) was distant from the site, making access via diving a lengthy prospect. Arranging adequate boat transport and support proved to be difficult, particularly because of scheduling constraints.
\end{enumerate}

\subsection{The Dives}
All dives (see Table~\ref{tab:dives}) were carried out by teams of divers using electronic closed-circuit rebreathers (eCCRs). Since eCCRs deliver breathing gas with a fixed partial pressure of oxygen (pO$_2$), they presented a considerable advantage when working at depth (50-60~m), enabling longer working duration while minimising the time required by the divers to decompress. The use of DPVs maximised the area that could be covered within a limited time-window. Each dive involved a minimum of two divers; one diver was fully dedicated to data acquisition, with the remaining divers being responsible for dive safety and/or video \& photo documentation of the work (see Fig.~\ref{fig:dpv_camera_setup}).  The site was accessed from the Inland Sea, and DPVs were used to arrive at the site and return to the exit point.

The duration of a typical, single dive was around 1.5 hours. The trip to the site and back entailed around 15 to 20 minutes either way, leaving around 45 minutes to 1 hour to set up and capture the data, which matches closely the maximum duration of the lights being used (on full power).

\subsection{Data acquisition}\label{sec:data_gathering}

\subsubsection{Image data}
\begin{figure}
  \includegraphics[width=\linewidth]{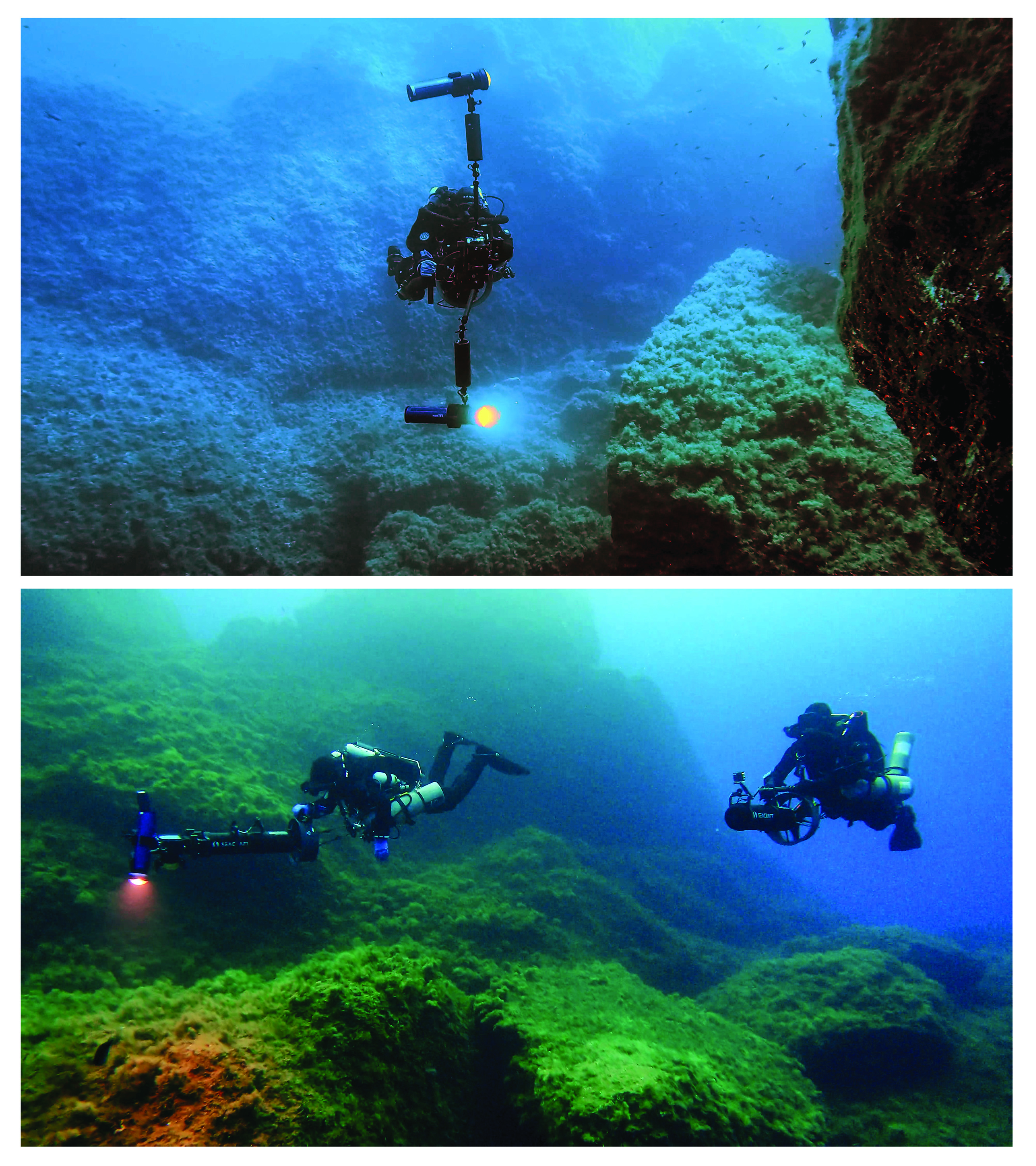}
  \caption{These two photographs show the setup that was used, with camera and lights assembled on a DPV. The diver responsible for photography was monitored at all times by a safety diver (bottom photograph). (Image copyright: J.~Caruana.)}
  \label{fig:dpv_camera_setup}
\end{figure}

Due to the size of the area to be surveyed, a handheld photography rig (consisting of camera and lights) controlled by a swimming diver was excessively time-consuming and tiring, and therefore not a viable approach. Therefore, both camera and lights were mounted on a DPV (see Fig.~\ref{fig:dpv_camera_setup}). 

\begin{figure}
  \includegraphics[width=\linewidth]{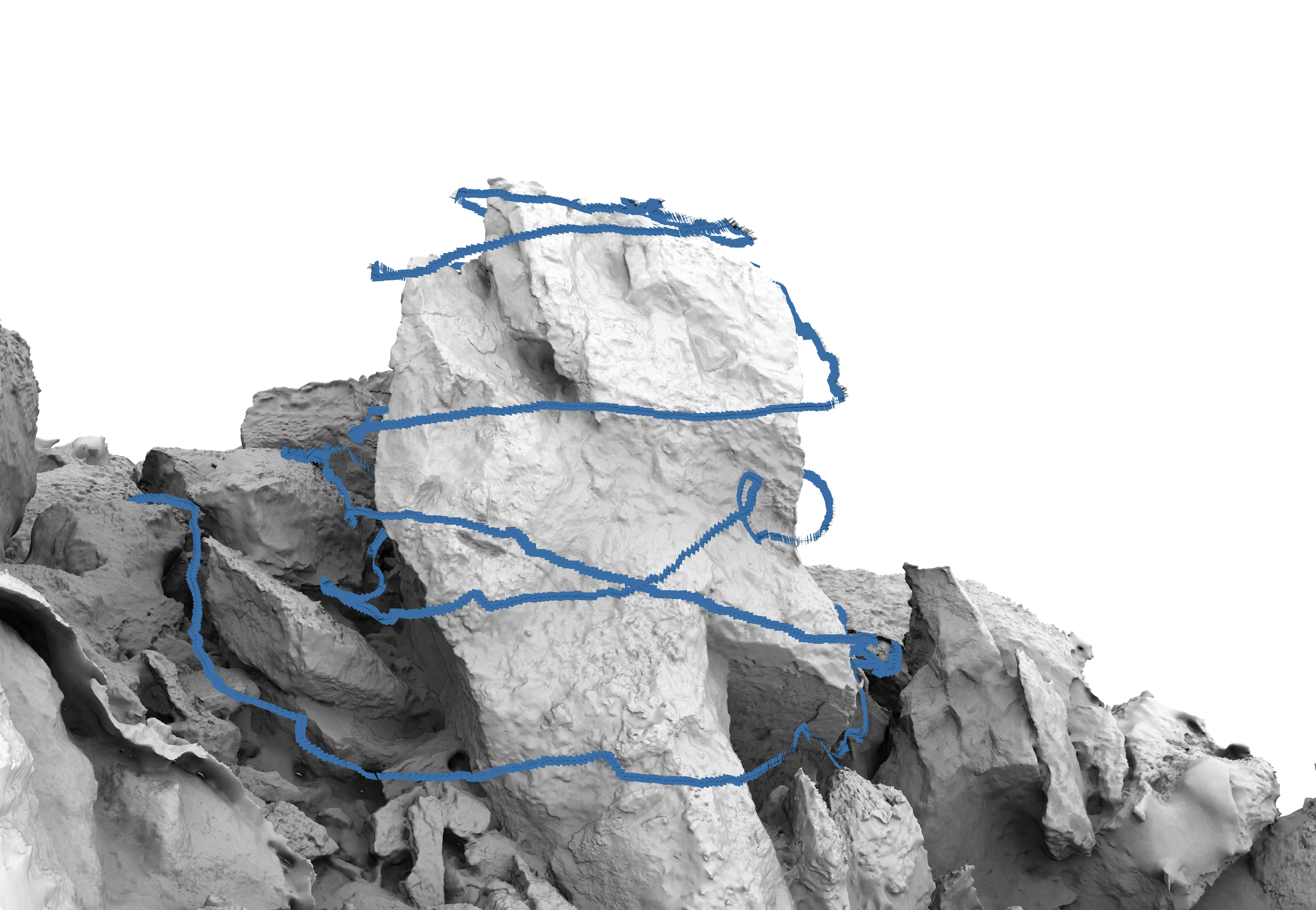}
  \caption{A single camera path around an elevated feature. In such cases, we adopted an approximately spiralling path, proceeding from the shallowest point to deeper waters.}
  \label{fig:camera_path}
\end{figure}

Once the diving team arrived at the site, the lights were deployed (by extending the support arms). On most occasions we used 2$\times$~30K~lm lights (Keldan); for two of the dives we used 3$\times$~12K~lm lights. Following their deployment, the diver capturing the data would execute the pre-determined path as closely as possible. When covering relatively flat terrain we followed a route that approximated a lawnmower path. For elevated features we adopted a spiralling path, proceeding from shallow to deeper water (as illustrated in Fig.~\ref{fig:camera_path}). On completion of the image acquisition (or when the lights' battery power was depleted), the lights would be re-stowed and the team head back together to the exit-point.

The main cameras used for data gathering were full-frame digital mirrorless cameras (Sony A7S and A7III). Photos were captured in both JPG and RAW format.

Camera exposure settings, specifically the shutter speed and aperture (f-number), were pre-calculated, while the ISO (sensitivity setting) was set to adjust to conditions automatically.  The shutter speed setting is especially important in such a use-case where one needs to find the right balance between adequately exposing the subject and limiting motion-blur (since the camera is mounted on a continuously moving rig). This was selected on the basis of the ground sample distance (GSD) and speed of movement through the water, calculated via:

\begin{equation}
t \leq \frac{{\rm GSD}}{v}
\end{equation}

where $t$ is the exposure time in seconds, $v$ the velocity in ms$^{-1}$ of the moving camera, and

\begin{equation}
{\rm GSD}=\frac{D}{f}\times {\rm pixel size} = \frac{D}{f}\times\frac{H_{\rm sensor (mm)}}{H_{\rm image (pixel)}}
\end{equation}

where $D$ the distance between camera and subject in mm, $f$ the system's focal length in mm, $H_{\rm sensor (mm)}$ the sensor height in~mm, and $H_{\rm image (pixel)}$ the image height in pixels.

The GSD, in mm, defines the smallest distinguishable detail of the object in a photo, or, similarly, the image pixel pitch projected in the object space. It depends on the size of the sensor, its resolution (number of pixels) and the image scale, i.e.~the ratio between the focal length of the camera used and the distance from the object at which the image was taken. The GSD achieved was in the range of 0.43-2.43~mm, far exceeding the requirements of the project.

The f-number was not as critical a parameter on account of the large depth of field resulting from the use of wide-angle lenses behind dome ports.  

Given the above considerations, typical values were as follows: a shutter speed of 1/250~s, an f-number of f/8, and an ISO ranging between 100 and 512,000.

A salient aspect was the choice of number of frames shot per second, aiming for sufficient image-area overlap to enable a successful photogrammetry process. We employed a shooting rate of 1 frame every 0.8~seconds, which yielded overlap with high redundancy; overlap in the direction of movement of 90-95\% was typical. We emphasise that these are theoretical and simplified calculations. In practice, external factors such as current, visibility and lighting conditions may be worse than expected, requiring adaptation of the acquisition planning on-the-flight. The lack of a real-time survey and verification tool can make it difficult to ensure that the acquisition is meeting the requirements (resolution and accuracy) even for experienced operators.

Since the number of photos obtained within a given time period was known (through the timestamp in EXIF data), and the distance travelled could be computed from the model, the approximate speed through the water could be derived; this amounted to $\approx$0.5~m~s$^{-1}$.

At all times, in addition to the principal camera, a GoPro was employed as a backup camera recording video. This video data was used only in one instance where some gaps in the data from the principal camera in use presented some difficulties in the photogrammetric image orientation. In this case, frames extracted from the GoPro camera videos were used to supplement the photographic data from the main camera.  A total of 525.3 (211.6) GB of raw (JPG) image data was captured across all dives.

While all dives rendered useful data, there was a steep learning curve during the initial dives which resulted in significant improvement in the approach as the project moved forward. During the first dive, which was intended to suggest a preliminary idea of the challenges involved, an area was selected at random for initial data capture. The next set of dives were aimed at gathering sufficient data to put together a coarse, low resolution 3D-model of the complete site, which could be utilised as a baseline model to better plan and execute subsequent data gathering. For this purpose we recorded video footage (instead of taking photos), operating at a larger distance from the subject, in order to cover a larger area in less time. In fact, this process was successfully completed over two dives (Dives \#2-3). Although the dataset thus acquired was not used for the final 3D-model, the baseline model we built from it proved invaluable to learn the topography of the site. Moreover, during later dives it helped us better understand which sections had been adequately captured and which ones remained to be covered. Consequently, the planning and execution of subsequent dives could be carried out more effectively.

\begin{sidewaystable}
\begin{center}
 {\tiny \begin{tabular}{| c | c | l | c | c | c | l | c | c | l |} 
 \hline
Dive & Date & Main Objective & Media & DPV? & Main & Lens & Lights & No. & Comments \\
 & & & Type & & Camera & & & frames & \\
 \hline\hline 

\rowcolor{white}
1 & 19/12/20 & [4] & Stills & Yes & A7S  & 16-35~mm & 3X Keldan 12K lm & 2,375 & Used scale bars. \\ 

\rowcolor{gray!25}
2 & 06/01/21 & Overall site & Video & Yes & A7S  & 16-35~mm & None & 6,005 & Stills extracted from \\ 
\rowcolor{gray!25}
 & & & & & & & & & video @ 0.5~f~s$^{-1}$.\\ 

\rowcolor{white}
3 & 19/01/21 & Overall site (detail) & Video & Yes & A7S  & 16-35~mm & None 
& 8,429 & Stills extracted from \\ 
\rowcolor{white}
 & & & & & & & & & video @ 0.5~f~s$^{-1}$.\\ 

\rowcolor{gray!25}
4 & 06/02/21 & [3]; S of [1]; [2]; [3] & Stills & No & A7S  & 16-35~mm & 3X Keldan 12K lm & 3,323 & Various GoPro photos \\ 
\rowcolor{gray!25}
 & & & & & & & & & used to bridge gaps.\\ 

\rowcolor{white}
5 & 02/03/21 & [1]; [3]; [4] (Minor: [2]; [7]) & Stills & Yes & A7III  & 12-24~mm & 2X Keldan 30K lm & 3,097 & Changed to A7III due to\\ 
\rowcolor{white}
 & & & & & & & & & exposure issues w/ A7S.\\ 

\rowcolor{gray!25}
6 & 06/03/21 & Top, N \& S of [1]; [2]; (Minor [3], [7]) & Stills & Yes & A7III  & 12-24~mm & 2X Keldan 30K lm & 3,419 & \\ 

\rowcolor{white}
7 & 08/03/21 & N \& S of [1]; W of [2]; [5] & Stills & Yes & A7III  & 12-24~mm & 2X Keldan 30K lm & 2,821 & \\  

\rowcolor{gray!25}
8 & 11/04/21 & N of [1]; [2]; [3]; W of [2]; 5 & Stills & Yes & A7III  & 12-24~mm & 2X Keldan 30K lm & 2,723 & \\ 

\rowcolor{white}
9 & 03/06/21 & Dedicated to measurements & & & & & & & Measurement tryout \\ [1ex] 

\rowcolor{gray!25}
10 & 21/07/21 & Dedicated to measurements & & Yes & A7III & 12-24~mm & 2X Keldan 30K lm & 2,045 & 7 measurement runs \\ [1ex]  

\rowcolor{white}
11 & 11/08/21 & S of [1]; W of [2]; [5]; [6]; [7] & Stills & Yes & A7III & 12-24~mm & 2X Keldan 30K lm & 1,696 & \\ 

\rowcolor{gray!25}
12 & 26/08/21 & S of [1]; W \& NW of [2]; [5]; [6] & Stills & Yes & A7III & 12-24~mm & 2X Keldan 30K lm & 2,201 & \\ 
[1ex] 

 \hline 

\end{tabular}}
\caption{A log of the dives that were carried out to obtain images and measurements. In the third column, numbers in square brackets refer to component numbers as labelled in Fig.~\ref{fig:ortho_labelled}, and letters indicate cardinal and intercardinal directions.}
\label{tab:dives}
\end{center}
\end{sidewaystable}

\begin{figure*}
  \includegraphics[width=\linewidth]{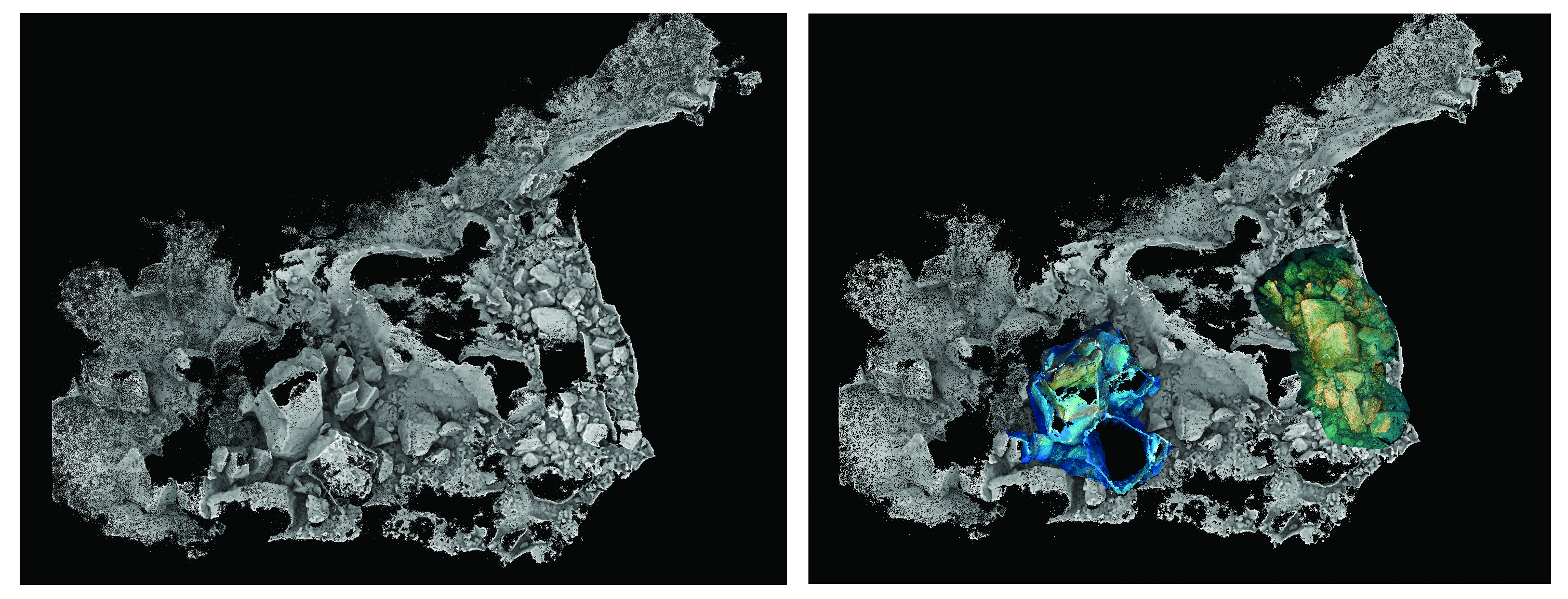}
  \caption{Left: The baseline model created from Dives \#2 and \#3. Right: the same baseline model with data from Dives \#1 and \#4 (shown in colour) integrated in it.}
  \label{fig:baseline_model}
\end{figure*}

As more dives were conducted, the gathered data was successfully merged with the baseline model built from dives \#2 and \#3 (see Fig.~\ref{fig:baseline_model}).  However, as we continued expanding the dataset, we achieved sufficient overlap between the high-resolution models such that the `baseline' model could eventually be discarded. From then on, each subsequent model was built by merging the latest dataset with the model that had been built from the previous dives, incrementally building the final model in stages. Each stage of completion provided the information necessary to plan subsequent dives.

\subsubsection{Scale and depth measurements}\label{sec:scaling}

First, we tried to measure distances on natural points with a metric tape. The visibility conditions, the morphology of the site and the presence of changing currents made these measurements difficult to acquire with an acceptable uncertainty for the purposes of the project. A different approach became necessary here. We therefore opted for the use of scale bars of 1~m and 2~m in length, placed in different parts of the site, and acquired during different dives. It is necessary to remark that this approach can provide an assessment of the local scale error, but not of possible global deformations of the photogrammetric model. The depth of seven well-recognisable natural points were measured with dive computers. This information, input into the photogrammetric model, was used to define the vertical reference system (datum).

\subsection{Processing}\label{sec:processing}

The processing stages\footnote{Readers interested in deepening their understanding of the theoretical aspects of the photogrammetric process may refer to \cite{luhmann2019} and \cite{forstner2016}.} for building the 3D-model involved the following basic steps:

\begin{enumerate}

\item Raw image preprocessing: the gathered RAW image data were converted in JPG file format with highest quality and minimum compression.

\item Image orientation and bundle adjustment: this step starts with the extraction and matching of homologous feature points in the images. Their 3D coordinates are then simultaneously calculated with the camera pose (position and attitude). Approximations of camera calibration parameters stored in the EXIF data are refined in the computation (self-calibration). This process is also known as structure from motion (SfM). Due to the complexity of the collected data, we adopted the following processing procedure. The images acquired in each individual dive were first processed separately, following a self-calibrating SfM procedure. The separate processes were then orientated together sequentially on the basis of overlapping areas. The orientation parameters estimated through this sequential process were used as an approximation to run a full SfM solution. This should ensure a more rigid connection between different acquisitions due to the fact that the features used for orientation are searched and matched on all images, including those from different acquisitions. The use of pre-calculated orientation parameters, meanwhile, helps to speed up the calculation. In this step, the camera calibration parameters were not refined further. At the completion of this process, the error on the scales and measured depths was 0.027~m and 0.29~m respectively.

\item Multi-view stereo (MVS) with mesh generation: Once the interior and exterior orientation parameters have been estimated, the next step is to generate a so-called dense model, i.e.~a model at the planned resolution for the current study. This model is generated through a procedure called multi-view stereo, in which homologous details of the object are searched and matched in all the images and reconstructed in 3D. There is a multitude of algorithms for solving this task \citep{nocerino2020}. In our case, we opted for the algorithm implemented in Agisoft Metashape, which provides the output of a polygonal model or mesh. The original mesh consisted of $10^9$ faces with an average resolution of 0.007~m \citep[calculated via the formula presented in][]{rodriguez2015}, which is much too high for practical purposes. We therefore generated a mesh with an average resolution of 0.058~m, which is sufficient to meet the requirements of this project.

\item Texturing: To improve the photorealistic rendering of the model, a further step is required, in which a biunivocal relationship is established between images and mesh geometry. In other words, portions of images are projected onto the polygons that constitute the mesh model and usually composed together in one or more images called external textures. The final image texture size was 8192$\times$8192~pixels$^2$, which was adequate for the aims of this project. Additionally, another product to enhance the visualisation and support the interpretation of the 3D model geometry was the ambient occlusion texture map. A grayscale value is associated to each surface point, coding the amount of diffuse light reaching the point itself. It can also be interpreted as the amount of sky visible from each surface point. The more occluded a point is, the darker the value.

\end{enumerate}

\section{Results and Discussion}\label{sec:geo_assessment}

\subsection{Seafloor morphology}

\begin{figure}[H]
  \includegraphics[width=\linewidth]{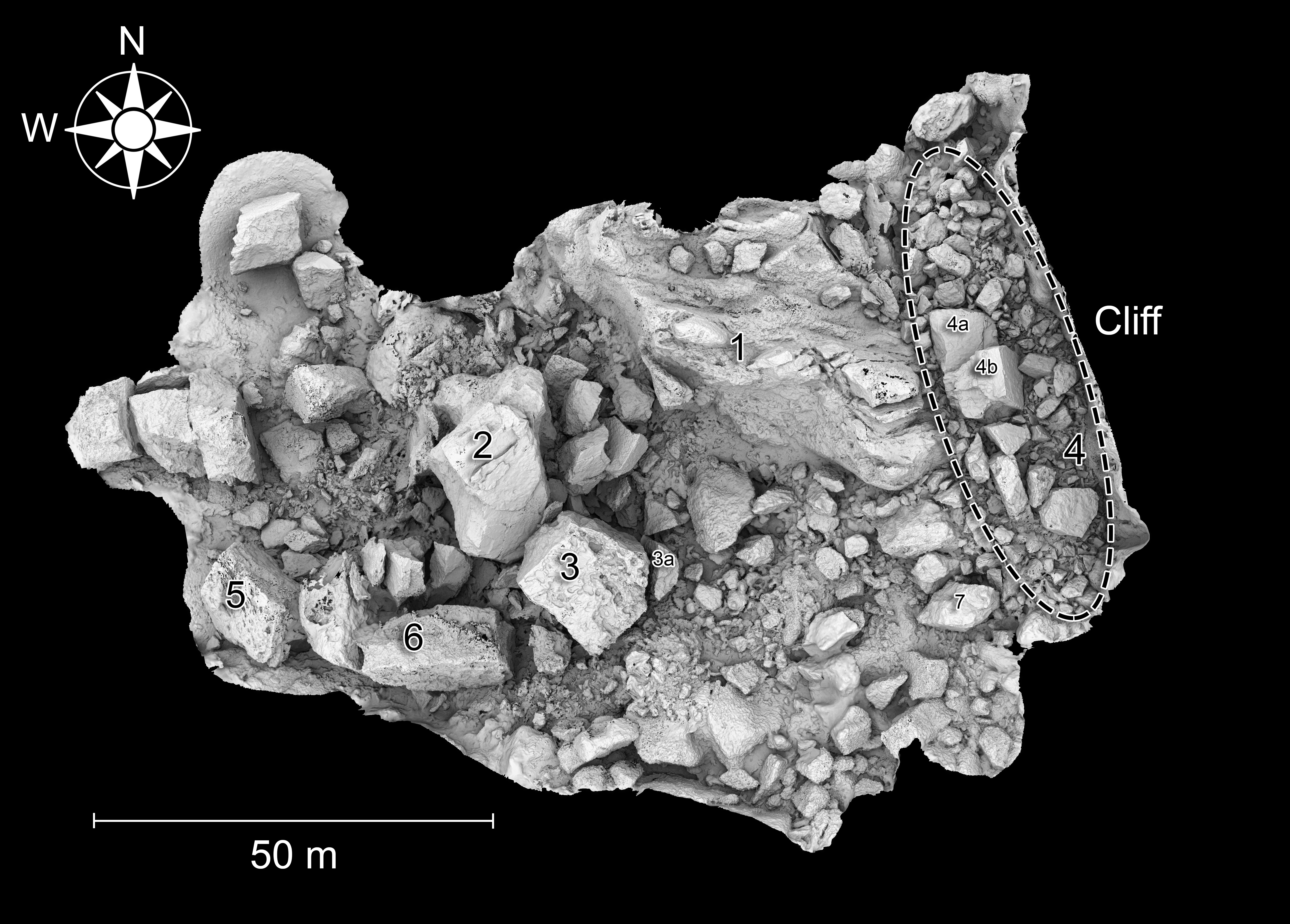}
  \caption{Ortho-image of the site of the former Azure Window generated with the software Metashape, xNormal, and Cloud Compare. Principal identified components are as follows: [1] Base of the pillar; [2] Attard member of pillar; [3] Component from Xlendi/Mara region atop pillar; [4] Remains of the Azure Window's bridge, with `4a' and `4b' being the bridge's two main components; [5] Component from Xlendi/Mara region atop pillar; [6] Component from Xlendi/Mara region atop pillar; [7] Attard member of pillar that collapsed in 2012. The vertical cliff to which the arch used to be connected is labelled on the right.}
  \label{fig:ortho_labelled}
\end{figure}

\begin{figure}[H]
  \includegraphics[width=\linewidth]{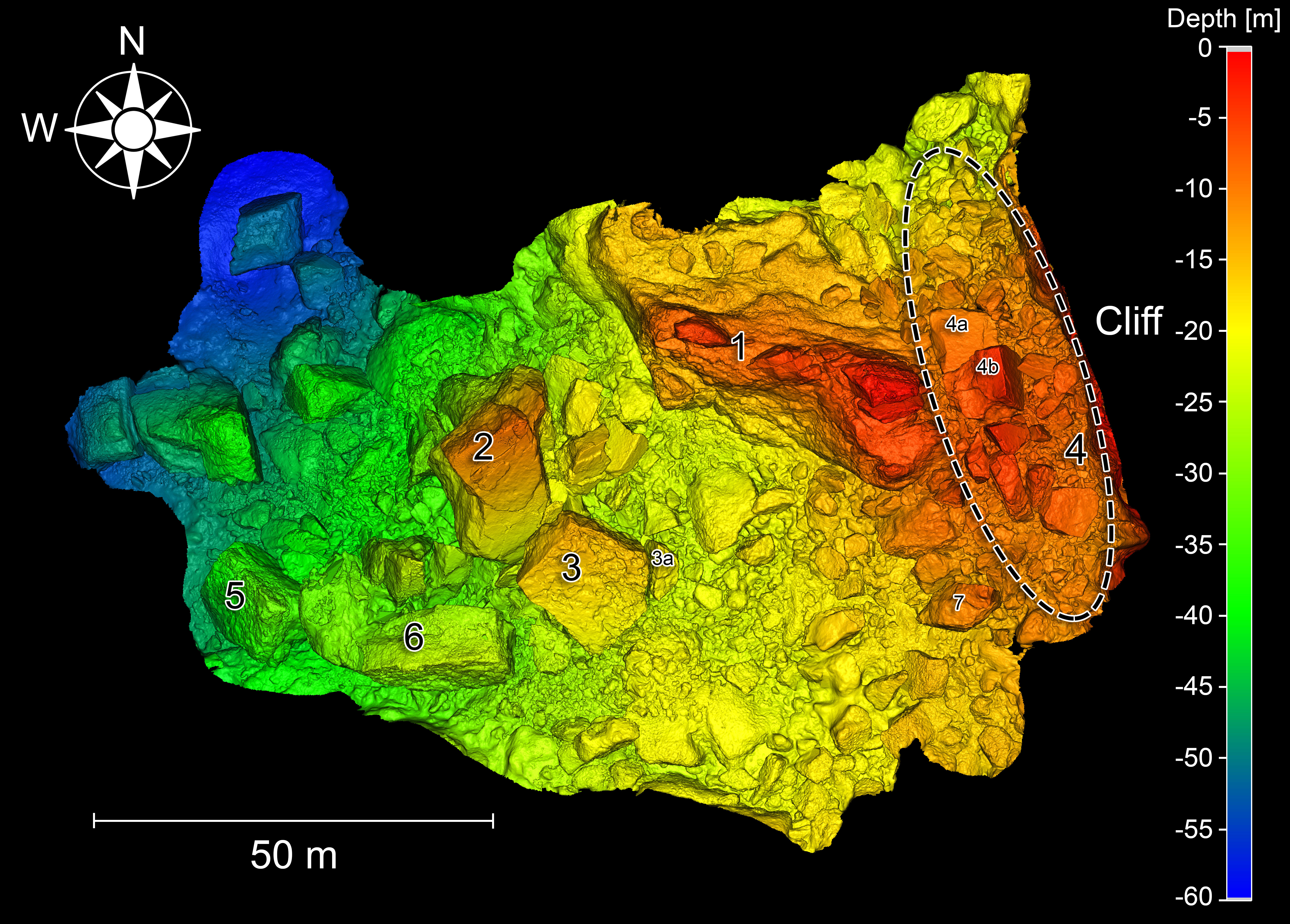}
  \caption{Depth-map of the site of the former Azure Window generated with the software Cloud Compare. Identified remains are scattered across a wide range of depths, from $\approx$3~m below the surface down to $\approx$60~m. The seabed slopes from shallow waters in the east to deeper waters in the west.}
  \label{fig:depth_map}
\end{figure}

An ortho-image and digital elevation model of the seafloor at the site of the collapse of the Azure Window are presented in Figures \ref{fig:ortho_labelled} and \ref{fig:depth_map}. 

From Fig.~\ref{fig:depth_map} one may appreciate how the seabed exhibits a slope from shallow waters in the east (beneath the cliff) to deeper waters in the west.

The following is a list of the main identified morphological elements and their approximate dimensions and depths:

\begin{enumerate}
\item [1.] 40~m long~$\times$~14~m wide (at base)~$\times$~15~m high ridge that tapers upwards towards the surface. The top of the ridge has a jagged surface with pointy sections, cracks, and steps which seem to follow the joint pattern of the rock. The northern flank has a regular notch at a depth of $\approx$8-9~m (see Fig.~\ref{fig:labelled_ridge}). Sections of the ridge in the north and northwest exhibit a stepped morphology, with the deepest point sitting at $\approx$28~m. The ridge extends to very shallow waters in some areas, with the shallowest section ($\approx$1.4~m) being located to the east.

\begin{figure}[H]
  \includegraphics[width=\linewidth]{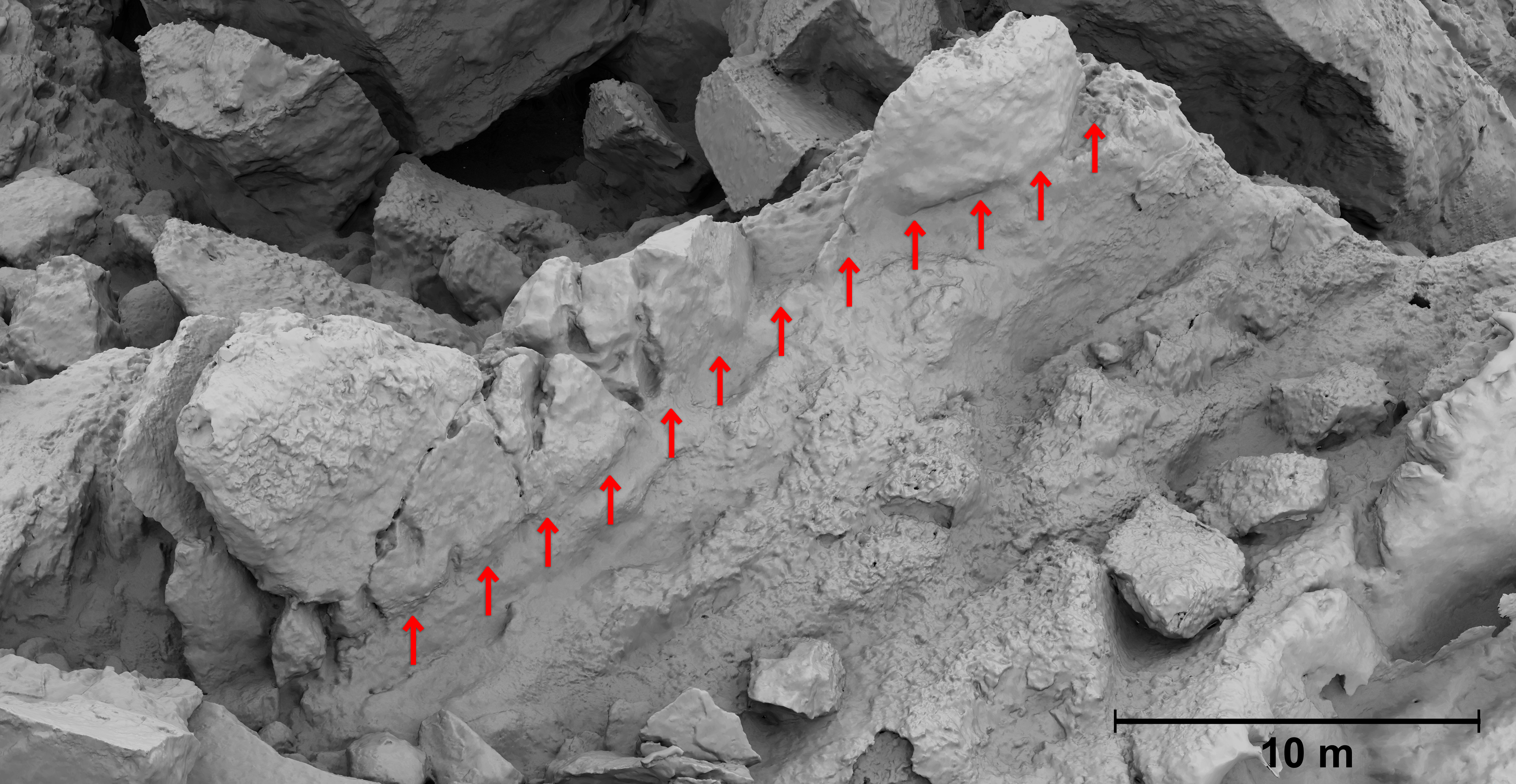}
  \caption{An ortho-image of Component 1, derived from the model using the software Metashape and Cloud Compare. The image shows the view when looking towards the southwest, exhibiting the northern face of the ridge (running diagonally from lower left to upper right of this image). The ridge, which tapers upwards towards the surface, exhibits a regular notch at a depth of $\approx$8-9m (marked with arrows) and several cracks and steps.}
  \label{fig:labelled_ridge}
\end{figure}

\item [2.] 14~m~$\times$~14~m$\times$~23~m (irregularly-shaped) Attard boulder, with Xlendi parts on the southeast-pointing face. The boulder is cuboidal at the base and tapers to a ridge at the top, where layered cracks follow the joint pattern. The southeastern face of the boulder is observed to be relatively planar, and parts of the northwestern face are fresh-looking. The deepest point sits at $\approx$36~m, and the shallowest at $\approx$9~m.

\item [3.] 14~m~$\times$~13~m~$\times$~8~m (roughly cuboid) boulder, with a weathered top and fresh-looking planar sides. The northeastern corner of the boulder has broken off as a triangular prism (labelled `3a'). The deepest point is at $\approx$24~m and the shallowest at $\approx$14~m.

\item [4.] 1,000~m$^2$ boulder field with rocks of various sizes, with a Mara boulder measuring up to $\approx$8~m wide. The majority of the boulders are less than $\approx$4~m wide and consist of triangular prisms or cuboids with fresh-looking planar sides. Some of the boulders taper upwards to form ridges. All boulders are located in relatively shallow waters ($\approx$4-23~m).

\item [5.] 10~m~$\times$~13~m~$\times~8$~m (at thickest part) cuboidal Xlendi/Mara boulder with a weathered southwestern face. The northwestern flank is relatively planar (and the same applies to the northeastern flank, albeit to a lesser degree). The boulder is located in deeper waters, with the deepest point at $\approx$46~m and the shallowest at $\approx$32~m.

\item [6.] 17~m~$\times$~8~m~$\times$~8~m weathered cuboidal boulder located in deeper waters, the deepest point being at $\approx$34~m and the shallowest at $\approx$24~m.
\end{enumerate}

\subsection{Discussion of the collapse of the Azure Window natural arch}

\begin{figure}[H]
  \includegraphics[width=\linewidth]{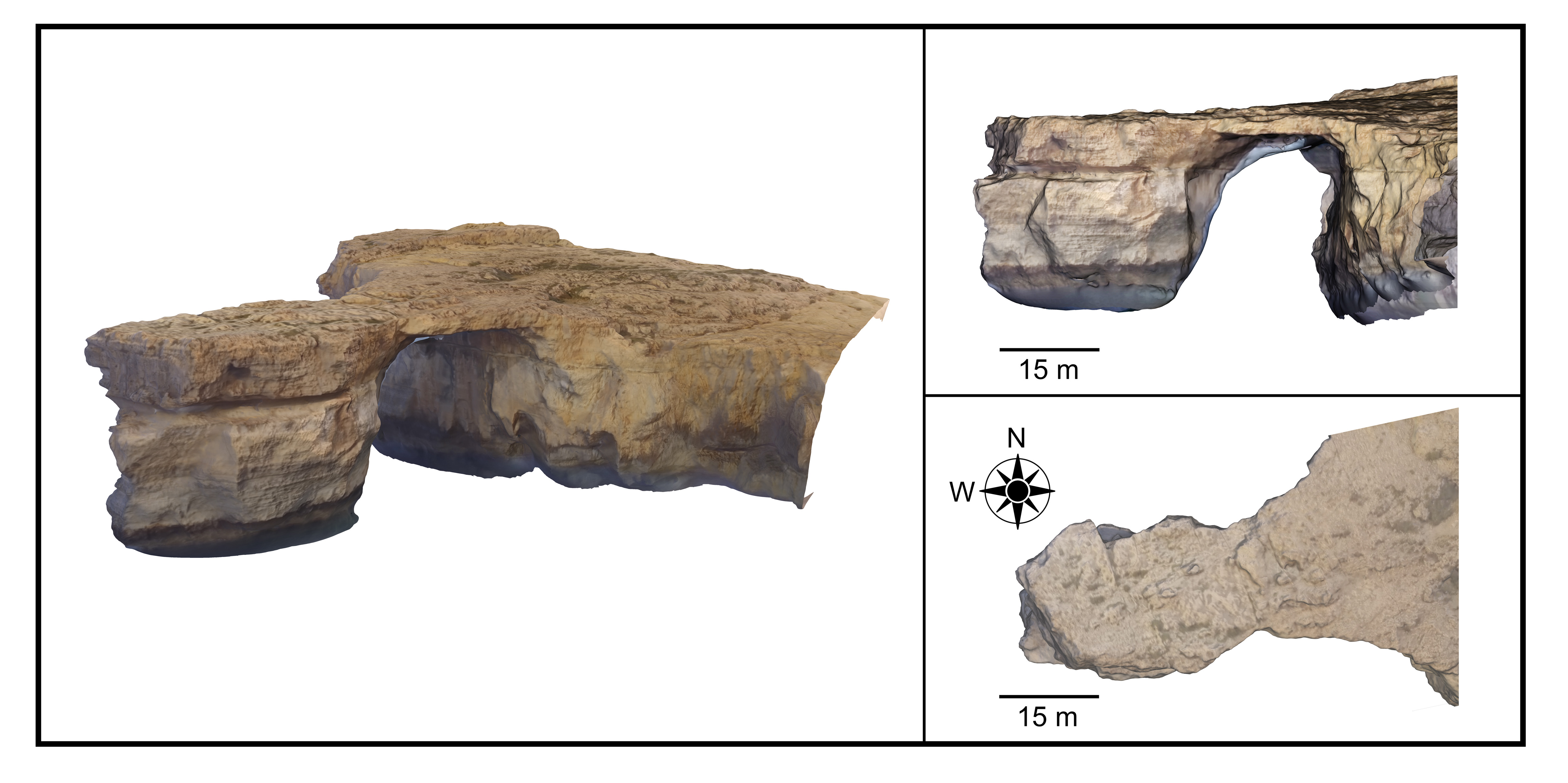}
  \caption{On the left, a 3D-model (created from drone footage) of the Azure Window prior to its collapse. On the right, two scaled ortho-images derived from the model using the software Metashape and Cloud Compare, showing the arch (top) and a bird's-eye view (bottom). (Note: the 3D-model on the left is not to scale with respect to the ortho-images on the right.)}
  \label{fig:pre_collapse_model}
\end{figure}

The interpretation that follows of the deposits and layout described above is based upon familiarity with the site via dives carried out prior to the collapse, a comparison of features against images and a 3D reconstruction of the Azure Window prior to its collapse (Fig.~\ref{fig:pre_collapse_model}), as well as observation of exposed rock segments in the submerged remains. 

\begin{figure*}
  \includegraphics[width=\linewidth]{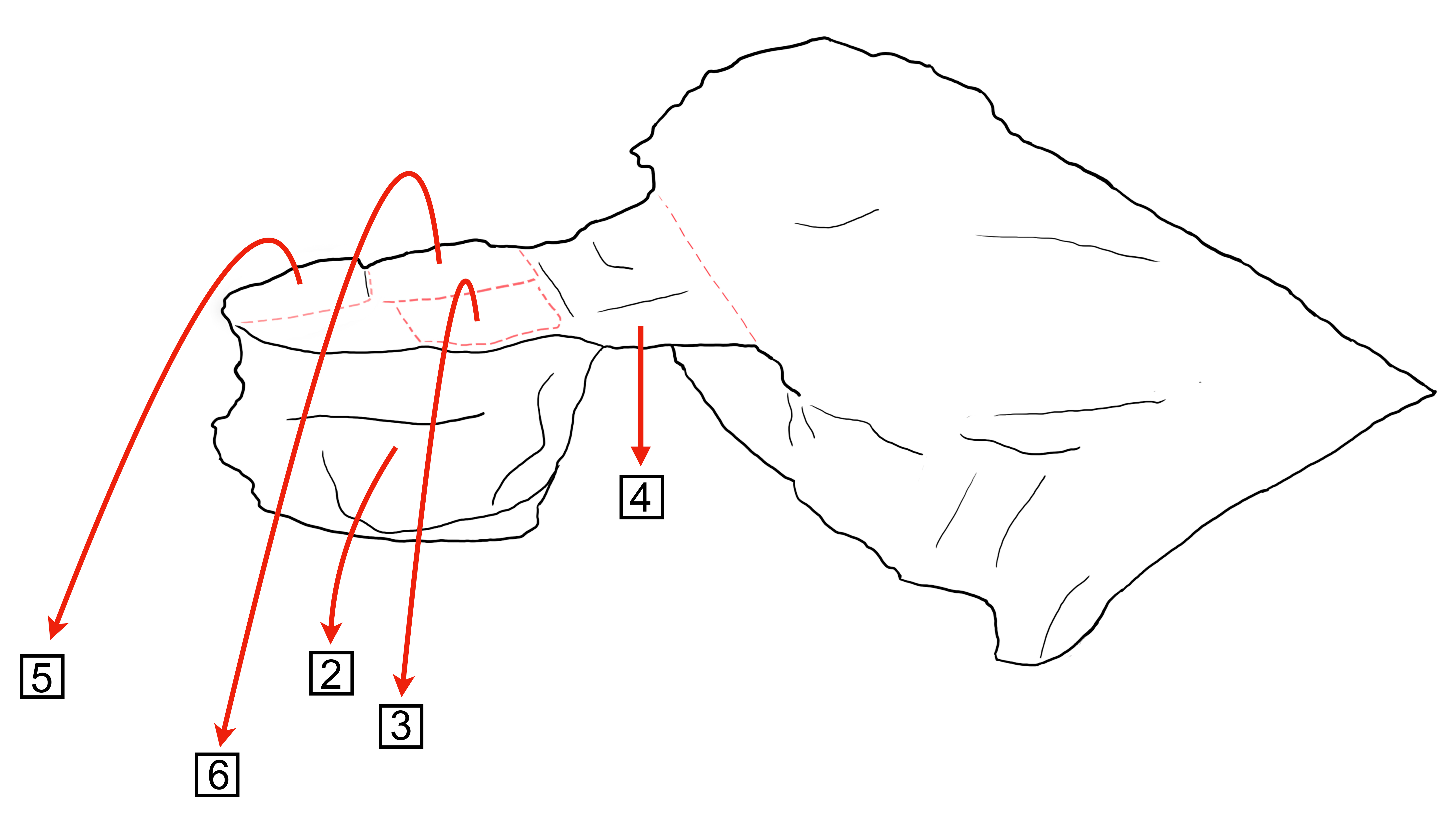}
  \caption{A simplified sketch that illustrates the final resting position of principal components of the Azure Window. Top sections (Xlendi/Mara) from the pillar [3, 5, 6] collapsed to the southwest, as did a large segment of the Attard member of the pillar [2]. The bridge [4] collapsed vertically upon loss of support from the pillar.}
  \label{fig:sketch_collapse}
\end{figure*}

Referring to Figs.~\ref{fig:ortho_labelled} and \ref{fig:sketch_collapse}, Component 1 is the base that supported the Azure Window's pillar. Component 2 in Fig.~\ref{fig:ortho_labelled} is part of the original pillar (Attard member, with Xlendi parts on the face that was previously in contact with the Xlendi component), whereas Components 3, 5, and 6 are sections from the top part (Xlendi/Mara) above the pillar. The boulder field (4) located right underneath the original location of the bridge consists of its fragmented remains.

\begin{figure*}
  \includegraphics[width=\linewidth]{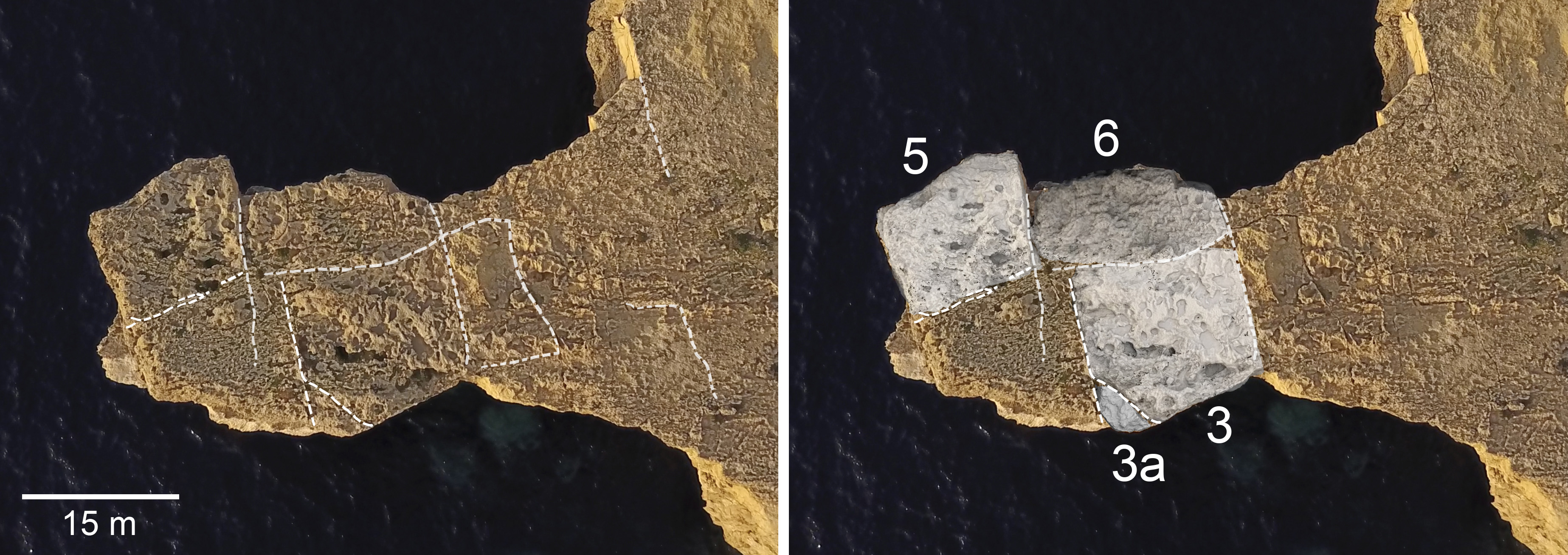}
  \caption{Surface lithology of the Azure Window. Dotted lines mark some prominently visible joints on the surface of the bridge. The right panel shows (in greyscale) three of the components extracted from the 3D-model of the underwater site overlain on the pre-collapse photo. It is clear that breakup of the Xlendi and Mara members atop the pillar followed the contour delineated by pre-existing jointing. (Image copyright: J.~Caruana.)}
  \label{fig:top_view}
\end{figure*}

We therefore reconstruct the collapse of the Azure Window as follows. The entirety of the pillar section above the waterline collapsed in the southwest direction, breaking into two principal components, `2' and `3' in Fig.~\ref{fig:ortho_labelled}, their shortest distance from the base of the pillar being $\approx$13~m and $\approx$17~m respectively. Other blocks of modest size and smaller fragments strewn in the vicinity of these large blocks represent other pieces from the collapse. The clean separation of components according to their type -  e.g.~`2' being Attard, and `3' and `5' being Xlendi/Mara - clearly indicate separation along the lithological boundary. It is also evident that separation of pillar segments followed pre-existing jointing; from Fig.~\ref{fig:top_view}, which shows three of the components (`3', `5', and `6') overlain on a pre-collapse aerial photo of the Azure Window, it can be appreciated that these fit perfectly within the outline formed by these joints. Some nearby chunks of rock from the collapse may either be slabs that failed before the pillar collapsed, or fragments that were formed following impact of the pillar with the seafloor. The final resting position of the principal components, wherein the most northerly sections (Components 5 and 6) ended up in the southernmost region of the seabed and vice-versa (see Fig.~\ref{fig:sketch_collapse}), corresponds to a scenario whereby the pillar toppled as one whole piece (towards the southwest), breaking up upon impact with the seafloor. 

The bridge (Component 4 in Fig.~\ref{fig:ortho_labelled}) collapsed vertically upon loss of support from the pillar, and broke into two main components (`4a' and `4b'), with the rest of it fragmenting into smaller pieces strewn across a $\sim$1,000~m$^2$ area (the boulder field marked `4' in Fig.~\ref{fig:ortho_labelled}). This breakup is consistent with the nature of the Xlendi and Mara members and their closely spaced jointing, as described in Sec.~\ref{sec:geosetting}.  The reconstructed series of events is corroborated by seismogram records \citep{galea2018}, which show a main energy arrival (collapse of pillar) followed by a longer duration, lower amplitude signal (collapse of bridge). 

\begin{figure*}
  \includegraphics[width=\linewidth]{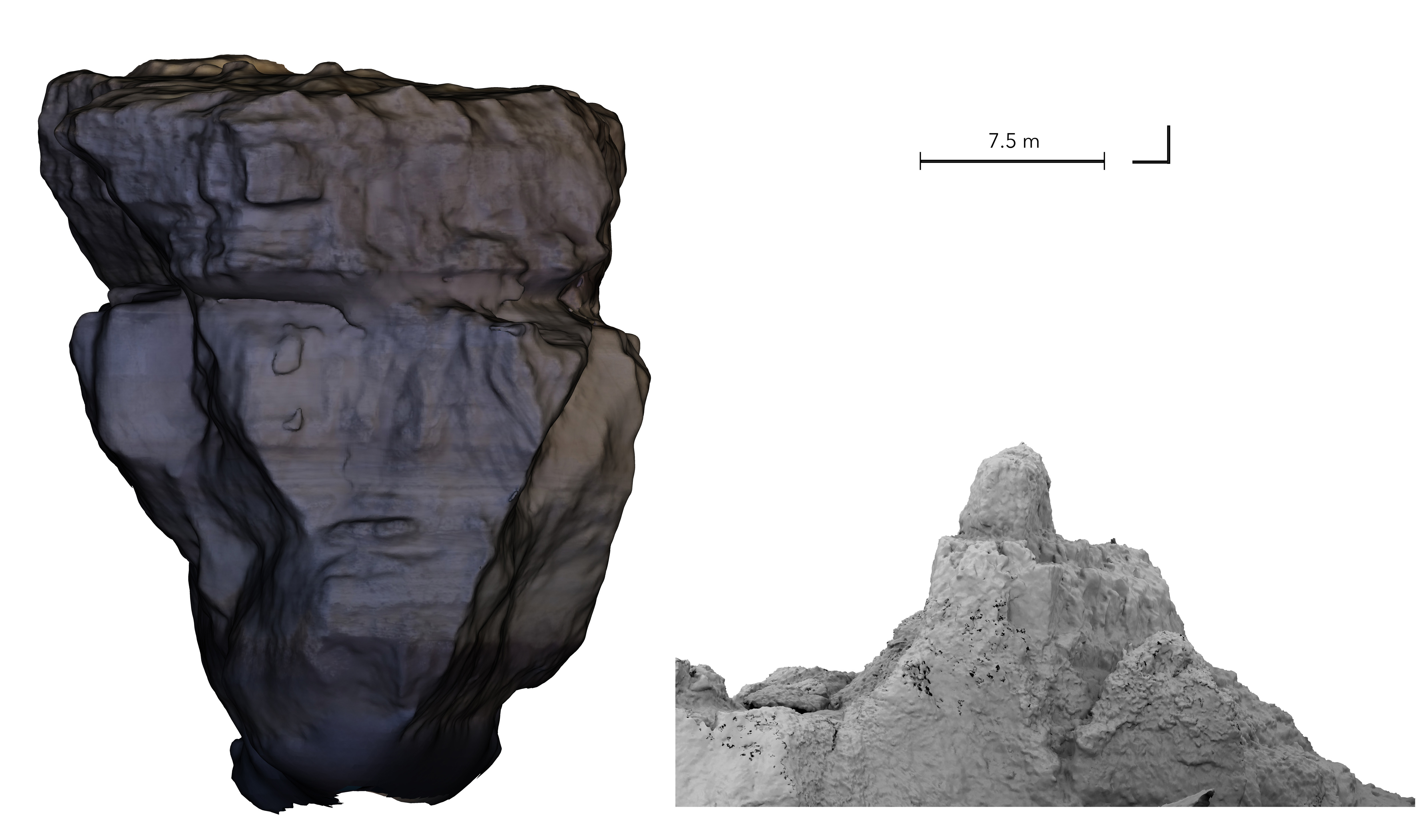}
  \caption{On the left, a side-view of the 3D-model of the (pre-collapse) Azure Window. On the right, a side-view (from the same angle) of the 3D-model of the (post-collapse) base that supported the pillar. (For both panels, the view is facing East.)}
  \label{fig:top_bottom_components}
\end{figure*}

The cause of the collapse of the pillar section is likely erosion at its base. The tapered ridge as well as a notch that was formed at the base of the pillar (Fig.~\ref{fig:labelled_ridge}) clearly point to progressive loss of support. (See Fig.~\ref{fig:top_bottom_components} for side-views of the pre-collapse Azure Window and the post-collapse base of the pillar.) The end-result was that no stack remained above the waterline. The collapse was precipitated by the violent storm that battered the site at the time, with numerical models of meteo-marine fields yielding wave-heights of up to around 3~m and wind-speeds larger than 16~ms$^{-1}$ \citep[see Fig.~5 in][]{galea2018}, in agreement with wind records that show wind gusts reached $\approx$18-20~ms$^{-1}$ before the collapse.

\subsection{Evolution of the site after collapse}

The most widespread change at the site since the collapse is rock surface erosion, which has significantly changed the visual appearance of the submerged remains, specifically their albedo; over time, the rocks' light, brightly-reflecting guise took on a duller aspect. 

Parts of the site have also experienced subsequent rockfall, leaving sections of the rock freshly-exposed, in turn facilitating continued erosion. This is most notable in the northwest-facing side of Component 2. We surmise that this side faces the brunt of hydraulic action on account of two factors: a northwesterly wind is predominant on the Maltese Islands, and this particular component is especially exposed since there is no rock of a significant size that can act as a barrier to the water flow and provide shielding.

\begin{figure*}
  \includegraphics[width=\linewidth]{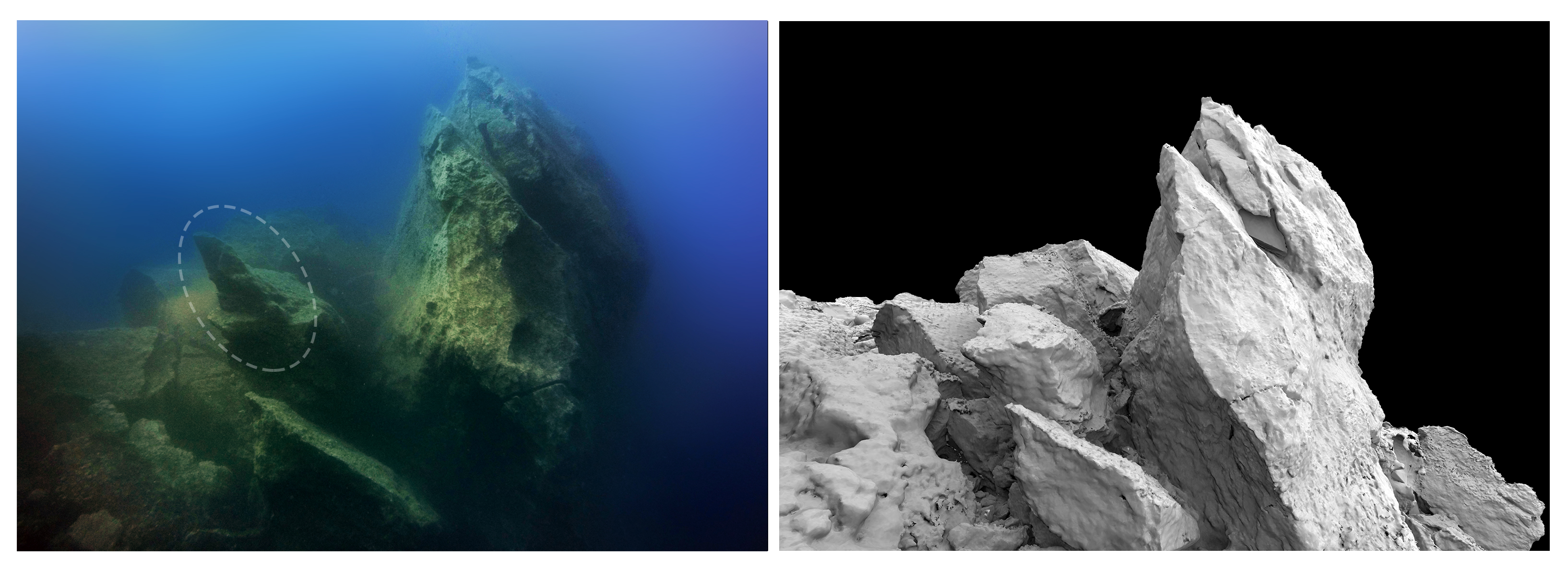}
  \caption{Comparison between a photograph acquired on the 22nd of August, 2018 and present-day photogrammetric representation of the same site, where the large component on the right is the Attard member of the pillar (i.e.~Component [2] in Fig.~\ref{fig:ortho_labelled}). Note the collapse of the feature circled in the photograph. (Left image is copyrighted by J.~Caruana.)}
  \label{fig:comparison_collapsed_piece}
\end{figure*}

Noteworthy instances of collapse have also been recorded, specifically in unstable/fragile rock formations that were particularly susceptible to destabilisation, as can be seen in Fig.~\ref{fig:comparison_collapsed_piece}. In this case, we also note that the formation happened to be completely exposed to the northwest, which probably exacerbated its condition. These observations lead us to anticipate that the morphology of some of the components might see further changes in the short-term.

\section{Conclusions}\label{conclusions}

In this paper we have demonstrated the successful application of moving-platform photogrammetry to a very-large-area underwater site with the aim of reconstructing the collapse event of a natural arch. 

The key stages in the collapse of the Azure Window involved erosion at the base of the pillar, which led to its collapse in the southwest direction, breaking into two principal sections with separation along the lithological boundary. We also find clear evidence that separation of some sections of the pillar followed pre-existing jointing. Upon losing support from the pillar, the bridge collapsed vertically, breaking into two main components and many other fragments.

Through this photogrammetric approach, corroborated with observation of exposed rock segments, we showed that it is possible to achieve identification of constituent components of the original structure even after significant time has elapsed and the rocks have undergone erosion and significant marine growth. This methodology can be effectively utilised to both understand and characterise similar events.

\section{Acknowledgements}

J.~Caruana acknowledges support from the University of Malta Research Support Services Directorate (PHYRP19-20 and PHYRP19-21). J.~Caruana and J.~Wood thank C.~Vella, D.~Marin, D.~Kovacevic, K.~Hyttinen, P.~Lammi, J.~Smith, and R.~Bartolo for assistance with the dives. J.~Caruana thanks B.~Azzopardi, G.~Mainente, and D.~Marin for very useful discussions.

\end{document}